\documentclass{sig-alternate}

\usepackage[pdftex,bookmarks=true,pdfborder={0 0 0},colorlinks=true,linkcolor=blue,citecolor=blue]{hyperref}
\usepackage{amsmath}
\usepackage{url}
\usepackage{subfigure}
\usepackage{booktabs}
\usepackage{array}
\usepackage{graphicx}
\usepackage{color}
\usepackage[table]{xcolor}
\usepackage{cite}
\usepackage{textcomp}
\usepackage{listings}
\usepackage{appendix}
\usepackage{tabularx}
\usepackage{rotating}
\usepackage{multirow}
\usepackage{amssymb}
\usepackage{amsfonts}
\usepackage{amsmath}
\usepackage{latexsym}
\usepackage{paralist}
\usepackage{util}

\newif\iftechreport
\techreporttrue
\newcommand\techreport[2][]{\iftechreport #2 \else #1 \fi}

\def\figurename{Figure}

\def\pvalue{\textit{p-value}}

\def\ds#1{\textsf{#1}}
\def\dsVerifiable{\ds{Verifiable}}
\def\dsVerified{\ds{Verified}}

\newcounter{researchquestion}

\newcommand{\researchquestion}[1] {
\refstepcounter{researchquestion}
\begin{description}
    \item [RQ\theresearchquestion] \emph{#1}
\end{description}
}

\newcounter{hypothesis}[researchquestion]
\renewcommand\thehypothesis{H\arabic{hypothesis}}

\newcommand{\hypothesis}[1] {
\refstepcounter{hypothesis}
\begin{description}
    \item [\thehypothesis] \emph{#1}
\end{description}
}

\newcounter{assumption}
\newcommand\assumptionname{ASS}
\renewcommand\theassumption{\assumptionname\arabic{assumption}}
\newcommand\assumption[1]{\refstepcounter{assumption}\theassumption #1}

\newenvironment{indentedlist}%
  {\begin{list}{}{\setlength{\labelwidth}{0pt}
   \setlength{\itemindent}{-\leftmargin}
   }}%
  {\end{list}}

\def\ver#1{v{#1}}

\def\cve#1{\textsc{#1}}

\def\WebKetRevision{80695}
\def\JSErevision{7138}

\def\TotalChromeNVD{\ensuremath{539}}
\def\TotalChromeConfirmedNVD{\ensuremath{503}}
\def\TotalCIT{\ensuremath{552}}
\def\TotalLocatedNVD{\ensuremath{167}} 
\def\TotalErrorLocatedNVD{\ensuremath{134}} 
\def\RunningTime{\ensuremath{16}}
\def\MachineRAM{\ensuremath{4}}
\def\MachineProcessor{3 x quad-core $2.83$GHz}

\def\Threshold{\ensuremath{5\%}}

\def\EErr{\emph{P-error}}
\def\LErr{\emph{F-error}}

\def\PErr{\emph{P-error}}
\def\FErr{\emph{F-error}}
\def\BErr{\emph{B-error}}

\begin{document}

\title{The (Un)Reliability of NVD Vulnerable Versions Data: an Empirical
Experiment on Google Chrome Vulnerabilities
\thanks{This work is supported by the European Commission under the project EU-SEC-CP-SECONOMICS}
}

\numberofauthors{1}
\author{
    \alignauthor Viet Hung Nguyen and Fabio Massacci \\
    \affaddr{DISI, University of Trento, Italy} \\
    \email{\{viethung.nguyen, fabio.massacci\}@unitn.it}
}

\maketitle

\begin{abstract}
NVD is one of the most popular databases used by researchers to conduct
empirical research on data sets of vulnerabilities. Our recent analysis on
Chrome vulnerability data reported by NVD has revealed an abnormally phenomenon
in the data where almost vulnerabilities were originated from the first
versions. This inspires our experiment to validate the reliability of the NVD
vulnerable version data. In this experiment, we verify for each version of
Chrome that NVD claims vulnerable is actually vulnerable. The experiment
revealed several errors in the vulnerability data of Chrome. Furthermore, we
have also analyzed how these errors might impact the conclusions of an
empirical study on foundational vulnerability. Our results show that different
conclusions could be obtained due to the data errors.
\end{abstract}


\section{Introduction} \label{sec:intro}

The last few years have seen a significant interest in empirical research on
data sets of vulnerabilities. Public third-party vulnerability databases, \eg
such as Bugtraq\cite{BUGTRAQ}, ISS/XForce{\cite{XFORCE}, National Vulnerability
Database (NVD)\cite{NVD}, Open Source Vulnerability Database
(OSVDB)\cite{OSVDB}, are mostly preferred by researchers due to their
diversity, availability, and popularity. Among these, NVD is one of the most
popular ones. The CVE-ID, \ie the identifier of each NVD entry, is usually used
as a common vulnerability identifier among other third-party data sources. In
this type of research, the quality of data sources play a crucial role in
empirical research on software vulnerabilities. If the data sources contain
wrong data, any conclusion derived from these data sources may be potentially
invalid.

Our research started from an abnormality in the data when we analyzed the NVD.
According to our analysis of NVD data, all of vulnerabilities in Chrome
\ver{2}--\ver{12} were originated from version \ver{1.0}. To explain this, the
following scenarios might occur: either yet more vulnerabilities in newer
versions have not been detected, or there is a problem in the vulnerability
data of Chrome, or both.

The analysis was based on an NVD data feature called \emph{`vulnerable software
and versions'} (or \emph{vulnerable versions} for short). This feature remarks
versions of particular applications that are vulnerable to the vulnerability
described in the entry. For example, \cve{CVE-2008-7294} lists all Chrome
versions before \ver{3.0.195.24} in its \emph{vulnerable versions}: this means
that vulnerability affects Chrome \ver{3.0} and all retrospective versions.
According to an archive document\footnote{This page is removed, but can be
accessed by url
\url{http://web.archive.org/web/20021201184650/http://icat.nist.gov/icat_documentation.htm}},
the information reported in this feature is \emph{``obtained from various
public and private sources. Much of this information is obtained (with
permission) from CERT, Security Focus and ISSX-Force"}. Furthermore, our
private communications with National Institute of Standards and Technology
(NIST), host of NVD, and software vendors, have revealed a ``paradox": NIST
claimed vulnerable versions were taken from software vendors; whereas, software
vendors claimed they did not know about this information. In other words, the
original source of this feature is unknown to the public, and therefore its
quality is unclear.

This raises a major threat to the validity of studies exploring this feature
such as
\cite{RESC-05-SP,OZMEN-SCHE-06-USENIX,YOUNIS-etal-11-SAM,MASS-etal-11-ESSOS,NGUY-MASS-12-ASIACCS},
and possibly others. We believe this may be a strong motivation to check for
the reliability of NVD.

\subsection{Contribution}
The major contributions of this work are as follows:
\begin{itemize}
    \item We present a replicable experiment to validate the reliability of
        ``vulnerable software and versions" feature of NVD for Chrome. This
        experiment can be applied for other open source applications (\eg
        Firefox, Linux).
    \item We show that the error rates of vulnerabilities in Chrome
        versions are significant. The errors are both erroneously reporting
        vulnerabilities in past and future versions.
\end{itemize}

The rest of the paper is organized as follows. We present our research question
and hypothesis (\secref{sec:question}). After that we describe the validation
method (\secref{sec:method}) that we follow to conduct the experiment. Next, we
report our result and perform analysis on collected data
(\secref{sec:analysis}). We also discuss the bias that might affect our studies
and how to mitigate them (\secref{sec:threat}). Next we briefly review studies
mostly related to our work (\secref{sec:relatedwork}). Finally, we conclude our
paper and discuss about the future work (\secref{sec:conclusion}).

\section{Running Example and Research Question}\label{sec:question}
We elaborate a running example on foundational vulnerabilities of Chrome to
study the impact of the (un)reliability of NVD data. A \emph{foundational
vulnerability} \cite{OZMEN-SCHE-06-USENIX} is one that was introduced in the
very first version of a software (\ie \ver{1.0}), but discovered later in newer
versions. In theory, foundational vulnerabilities have higher chance to be
exploited than others because they are exposed to attack longer than others. By
finding these vulnerabilities in \ver{1.0}, attackers could use them to exploit
recent versions (say, \ver{20}) at the release date. As the result,
foundational vulnerabilities are a source for zero-day exploits.

By June 2012, NVD reported $539$ vulnerabilities for $12$ stable
versions\footnote{\url{http://omahaproxy.appspot.com/about}, visit on July
2012. This is a web application supported by Google team for tracking
releases.} of Chrome\footnote{We only consider 1+ year old versions as to allow
their vulnerability data to mature.}. Out of these, $460$ ($85.3\%$) are
reported as foundational. \figref{fig:vulns:chrome} depicts the fraction of
foundational vulnerabilities of Chrome. Clearly, each analyzed version of
Chrome is rife with foundational vulnerabilities: $99.5\%$ on average are
foundational. We find unlikely that Chrome developers introduced a lot of
vulnerabilities in the first version, but none was introduced for the
subsequent $11$ versions. This motivate our research question  as follows.

\researchquestion{To what extent is the `vulnerable versions' feature of the
data reported by NVD truthworthy?}

To have such knowledge, for each pair of NVD entry and software version listed
in the \emph{vulnerable versions} data feature, we verify whether the NVD entry
impacts the corresponding version or not. If it is not, this pair is an
\emph{error}. The ratio of the number of error and the number of pairs is the
\emph{error rate} which we use as an indicator for the unreliability. In many
cases, a small error rate is acceptable. Depending on the type of study the
acceptable threshold of errors may vary. Here we choose the threshold of
\Threshold\ which is normally considered a threshold for statistical
significance. We consider the error rate as significant if the median of error
rates in individual versions is significantly greater than \Threshold. We test
the median of error rates, rather than the mean because a previous
study\cite{MASS-etal-11-ESSOS} has shown that vulnerabilities do not follow the
normal distribution. Hence, we test the following hypothesis:

\hypothesis{The median of error rates for vulnerabilities reported in Chrome
versions is greater than \Threshold.}\label{hypo:significant:err-rate}

We employ the non-parametric tests for the median to check for the
significance.

\begin{figure}
    \centering
        \includegraphics[width=0.7\columnwidth]{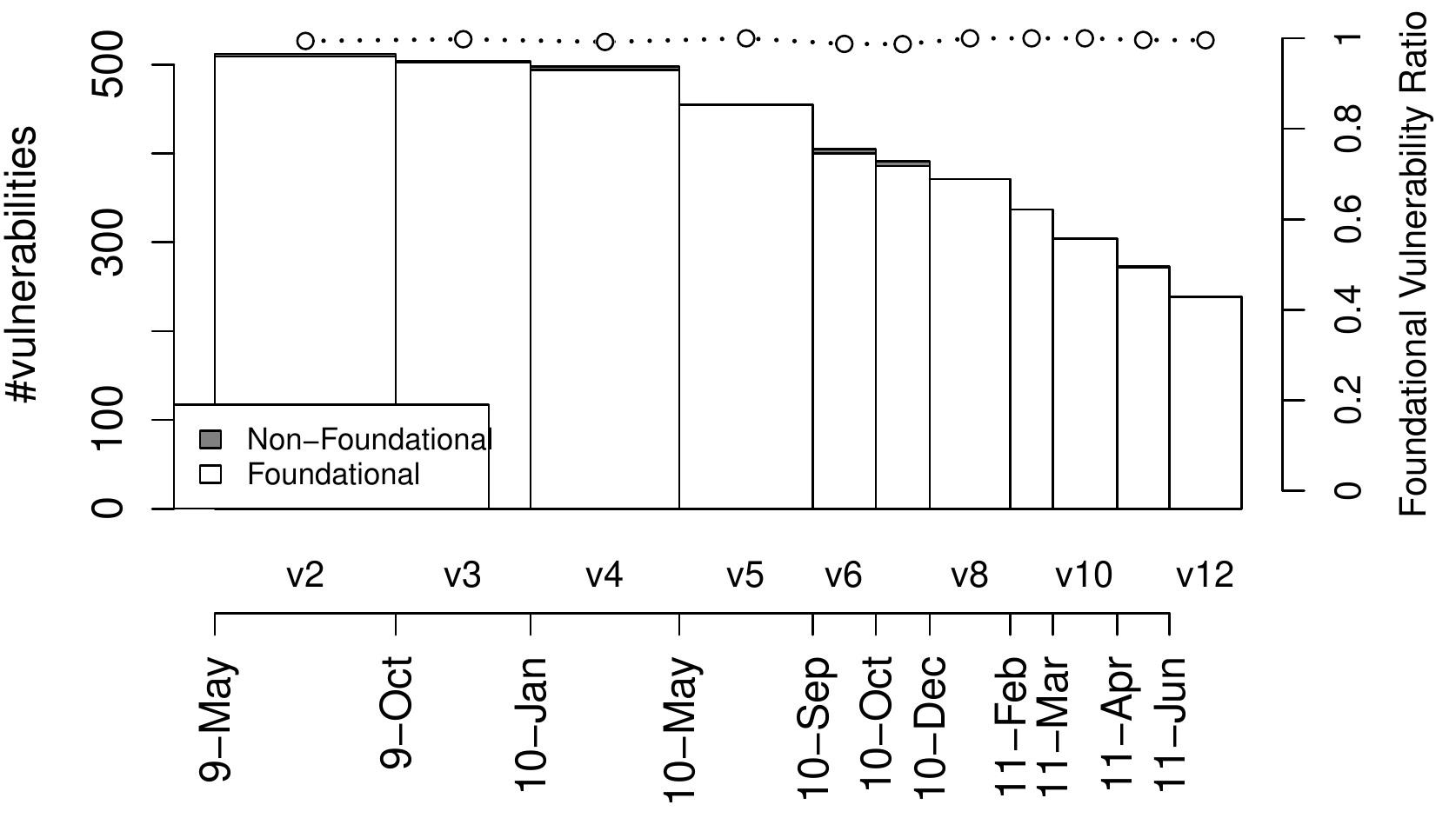}
    \ACCORCIA[-1]
    \extcaption[0.48\textwidth]{Above it is the stack bars represent the fraction of foundational
    and non-foundational vulnerabilities, below it is the release date of the versions. }
    \ACCORCIA
    \caption{Foundational vulnerabilities in Chrome.}
    \label{fig:vulns:chrome}
\end{figure}


\section{Validation Method} \label{sec:method}
To verify \TotalChromeNVD\ vulnerabilities for $12$ versions of Chrome, we need
to check $5,158$ pairs of vulnerability and version. Such huge amount of pairs
is impossible for a manual verification. Additionally, the manual approach is
 not replicable. Thus, our proposed method is based on a repeatable
and automatic approach \cite{SLIW-05-MSR} where security bug fixes are traced
back to the code base to locate the vulnerable code responsible for
vulnerabilities. Then we can determine whether a version claimed as vulnerable
is actually vulnerable. Our method relies on the following assumptions.

\begin{description}
    \item [\assumption \label{ass:commit:message}] When developers commit a
        bug-fix, they denote the bug ID in the commit message.
    \item [\assumption \label{ass:responsible:code}]If the fragment of code
        responsible for a vulnerability is not there then the software is
        not vulnerable.
    \item [\assumption \label{ass:responsible:missing}]If a vulnerability
        is fixed by only adding code to a vulnerable file, all versions
        containing the non-fixed revision of the vulnerable file are
        vulnerable.
\end{description}

By \emph{vulnerable files} we mean the files developers changed to fix a
vulnerability. Also, by \emph{code responsible} for a vulnerability we mean the
code that developers changed to fix the vulnerability. In some cases, the
changed code might not be the vulnerable one, but it helped to remove the
vulnerability, despite the original buggy code not being edited. For example, a
vulnerability that could lead to SQL injection attack could be fixed by
inserting a sanitizer around the source in another module. However, missing of
such sanitizer does not mean the application is vulnerable to the same SQL
injection attack. Even though the changed code is not buggy in some cases, we
still abuse the concept and call it \emph{code responsible}.

\begin{figure}[t]
    \centering
    \includegraphics[width=1\columnwidth, height=12\baselineskip]{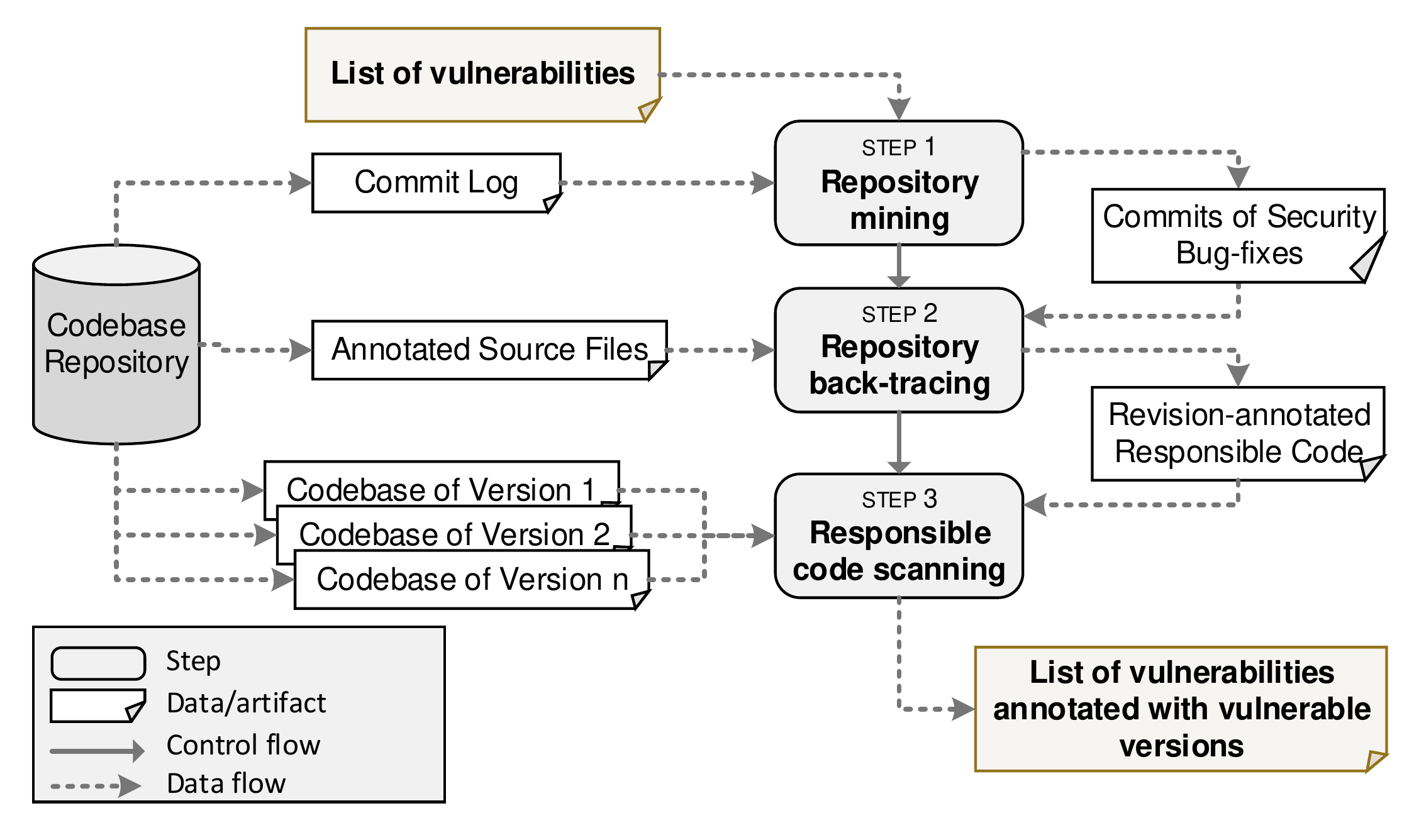}
    \ACCORCIA[-2]
    \caption{Validation method overview.}
    \label{fig:activities}
    \ACCORCIA
\end{figure}

\figref{fig:activities} sketches the steps of the proposed method. The input of
the process is a list of vulnerabilities and the output is list of
vulnerabilities annotated with vulnerable versions. The details are as follows:

\begin{indentedlist}
    \item [\sc step 1] \emph{Repository mining}. This step takes the list
        of vulnerabilities and the commit log (\ie list of commits)
        generated by the repository to produce commits of security bug
        fixes.  A commit of security bug-fixes is one that mentions a
        security bug ID\footnote{Bugs appear in the input list of
        vulnerabilities.} in its commit message and in some special
        patterns. These patterns may be vary in different software. For
        Chrome, they are \texttt{BUG=n(,n)*}, or
        \texttt{BUG=http://crbug.com/n} where $n$
        is the bug ID. 


    \item [\sc step 2] \emph{Repository back-tracing} This step takes
        commits of security bug-fixes, and the annotated source files from
        the repository to produce revision-annotated responsible lines of
        code (LoC). For each source file $f$ in each commit, let
        $r_{fixed}$ be the revision of this commit. We compare revision
        $r_{fixed}$ to revision $r_{fixed}-1$ of file $f$ using the
        \texttt{diff} command supported by the repository. The comparison
        output is in Unify Diff format, as exemplified by
        \figref{fig:snapshot:diff}, where we compare revision $r95730$ and
        $r95731$ of file \texttt{url\_fixer\_upper.cc}. By definition,
        responsible LoC appears in $r_{fixed}-1$, but not in $r_{fixed}$.
        For instance, from \figref{fig:snapshot:diff}, the responsible LoC
        is \Set{542}. We ignore trivial responsible LoC such as empty
        lines, or lines that contain only '\{' or '\}'.

        Next, we execute \texttt{annotate} command for $r_{fixed}-1$ of
        file $f$ to obtain the revisions of responsible LoC.
        \figref{fig:snapshot:annotate} presents an excerpt of the annotated
        file \texttt{url\_fixer} \texttt{\_upper.cc}. We see that the
        revision of LoC 542 is $r15$.

        There is a special case where the comparison between $r_{fixed}$
        and $r_{fixed}-1$ contains no line preceded with the minus sign. It
        means developers fixed the vulnerability by adding code only (\eg
        security check). In this case we assume that all versions
        containing revision $r_{fixed}-1$ and lower are vulnerable (see
        \ref{ass:responsible:missing}).

    \item [\sc step 3] \emph{Responsible code scanning.} This step looks
        for each revision-annotated responsible LoC in the code base of
        every version. If found, we append the corresponding version and
        the LoC to a list of version-annotated responsible LoC (see
        \ref{ass:responsible:code}). From this list, we can identify
        vulnerable versions for each vulnerability.
%
%
\end{indentedlist}

\begin{figure}[t]
    \centering
    \includegraphics[width=1\columnwidth,height=7\baselineskip]{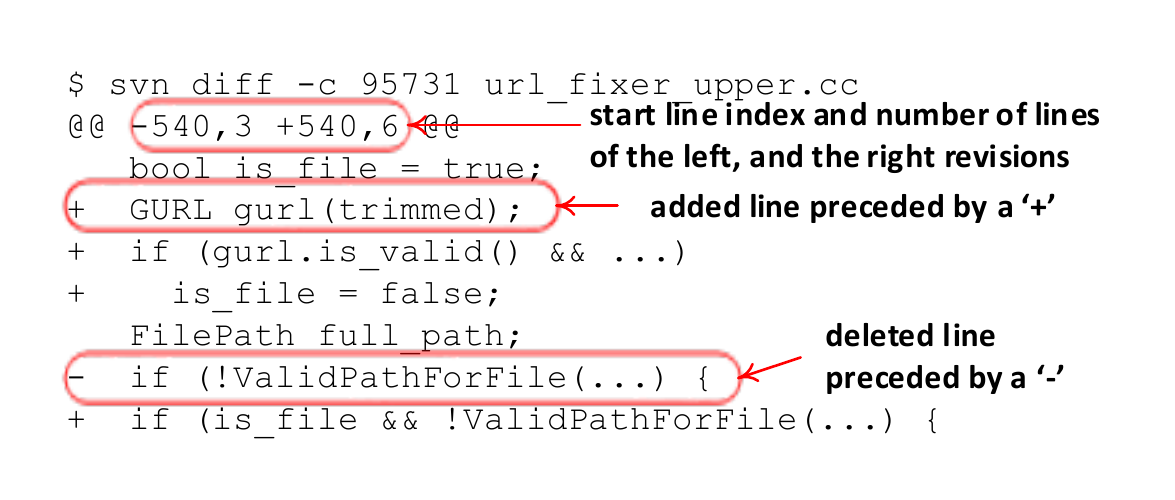}
    \ACCORCIA[-2.5]
    \caption{An excerpt of the \texttt{diff} of two revisions.}
    \label{fig:snapshot:diff}
    \ACCORCIA
\end{figure}

\begin{figure}[t]
    \centering
    \includegraphics[width=1\columnwidth,height=7\baselineskip]{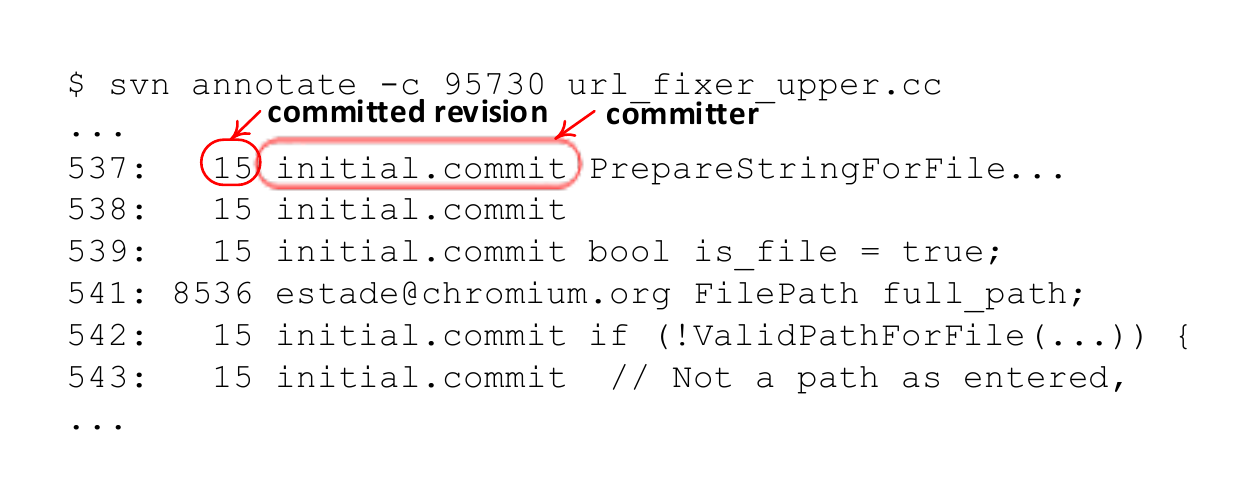}
    \ACCORCIA[-2.5]
    \caption{An excerpt of the annotation.}
    \label{fig:snapshot:annotate}
    \ACCORCIA
\end{figure}

\begin{table*}
  \centering \scriptsize \caption{The execution of the method on few Chrome vulnerabilities.}
    \vspace{5pt}
    \resizebox{0.9\textwidth}{!}{
    \begin{tabular}{C{23ex}C{18ex}cC{33ex}C{35ex}C{22ex}}
    \toprule
    \multicolumn{2}{c}{\sc input} &Output of  \sc step 1 & Output of \sc step 2 & \multicolumn{2}{c}{Output of \sc step 3} \\
    \cmidrule(r){1-2} \cmidrule(r){3-3} \cmidrule(r){4-4} \cmidrule(r){5-6}
    CVE-ID of NVD entry& Corresponding Bug & Commits for Bug-fixes & Revision-annotated responsible LoC & Version-annotated responsible LoC & Verified vulnerable versions \\
    \midrule
    2011-2822 (\ver{1}--\ver{13}) & 72492 & url\_fixer\_upper.cc$^1$ ($r95731$)& \Seq{r15,542} &  \Seq{\ver{1}-\ver{13}, 542}   & \ver{1}--\ver{13} \\
    2011-4080 (\ver{1}--\ver{8}) & 68115 & media\_bench.cc$^2$ ($r70413$) & \Seq{r26072, 352}, \Seq{r53193,353}& \Seq{\ver{3}-\ver{8}, 352}, \Seq{\ver{5}-\ver{8}, 353} & \ver{3}--\ver{8} \\
    2012-1521 (\ver{1}--\ver{18}) & 117110 & --  & --  & --  & -- \\
    \bottomrule
    \end{tabular}}
    \ACCORCIA \extcaption[0.9\textwidth]{$^1$:
    \texttt{chrome/browser/net/url\_fixer\_upper.cc} \hspace{5pt} $^2$:
    \texttt{media/tools/media\_bench/media\_bench.cc}}
    \label{tbl:method:example}%
    \ACCORCIA[-2]
\end{table*}%

Notice that there are \emph{unverifiable} vulnerabilities for which the method
can not verify the corresponding vulnerable versions. This could be due to a
couple of reasons. First, there is no corresponding security bug for a
vulnerability. Second, {\sc step 1} may not be always able to determine the
commit for security bug-fixes of vulnerabilities.

\tabref{tbl:method:example} shows a few examples of Chrome vulnerabilities
where we apply the method to verify their vulnerable versions. The two first
columns indicate the input of the method where we have list of NVD entries and
and their corresponding bug. We additionally annotate the vulnerable versions
reported by NVD for each entry next to the CVE-ID. The next columns show the
outputs of the steps. The dash line indicates the data is not available. It
means the corresponding NVD entry is not verifiable.

For a better understanding, we describe how the NVD entry \textsf{2011-2822} is
verified as in \tabref{tbl:method:example}. This vulnerability is reported to
affect Chrome \ver{1} up to \ver{13}. Its corresponding bug is \textsf{72492}.
In {\sc step 1}, by scanning the log, the bug fix is found at revision $r95731$
of file \texttt{url\_fixer\_upper.cc}. In {\sc step 2}, we diff revision
$r95730$ and $r95731$ of this file (see \figref{fig:snapshot:diff}). The
responsible LoC is determined as \Set{542}. Then we annotate $r95730$ of the
file to get the revision of the responsible LoC, which is \Set{r15} (see
\figref{fig:snapshot:annotate}). In {\sc step 3}, we scan for this line in the
code base of all versions, and found it in \ver{1} to \ver{13}. Finally, we
identify the vulnerable versions for this vulnerability, which are
\ver{1}--\ver{13}.


DISCUSS ABOUT MANUAL VERIFICATION OF THE RESULT

\section{Results and Example Revised} \label{sec:analysis}
We apply the proposed method to verify vulnerabilities of major versions of
Chrome from \ver{1} to \ver{12}. By June 2012, NVD reported \TotalChromeNVD\
entries\footnote{Observation on July 2012} that allegedly affect these versions
of Chrome. Out of these, \TotalChromeConfirmedNVD\ entries have links to
\TotalCIT\ security bugs in Chrome Issue Tracker in their \emph{references}
section. The method took \RunningTime\ hours on a \MachineProcessor\ Linux
machine with \MachineRAM GB of RAM to complete.

As the result, \TotalLocatedNVD\ NVD entries ($31\%$) are \emph{verifiable},
and $372\;(69\%)$ are unverifiable.  Among the verifiable ones,
\TotalErrorLocatedNVD\ $(81\%)$ have errors, \ie their verified vulnerable
versions are different than reported ones. Among the unverifiable, $36 (10\%)$
do not have corresponding bugs, and for $336 (90\%)$  we could not locate their
commits of security bug fixes. We have done a qualitative analysis on these
entries and found that they are bugs in external projects used in Chrome, \eg
WebKit -- the HTML rendering engine, V8 -- the java script engine, and so on.
Therefore, their commits of bug fixes do not exists in the repository of
Chrome. Later we will discuss how to work around this problem as a part of
future work.

\def\vuln{\textit{cve}}
\def\Vreport{\ensuremath{V}}
\def\Vverified{\ensuremath{V'}}
\def\version{\ensuremath{v}}
\def\verified{\textit{verified}}
\def\erroneous{\textit{erroneous}}
\def\unverifiable{\textit{unverifiable}}

In the following, we analyze the difference of vulnerabilities in individual
Chrome versions. Let \vuln\ be an NVD entry, and \version\ be the version in
analyzed, we define:
\begin{itemize}
    \item \Vreport(\vuln): is a set of reported vulnerable versions of
        \vuln.
    \item \Vverified(\vuln): is a set of verified vulnerable versions of
        \vuln. If \vuln\ is unverifiable, $\Vverified(\vuln) = \bot$.
    \item $\verified(\version) = \Set{\vuln | \version \in
        \Vverified(\vuln)}$: is the \vuln\ which responsible code is
        detected in version \version.
    \item $\erroneous(\version) = \Set{\vuln| \version \in \Vreport(\vuln)
        \wedge \version \not\in\Vverified(\vuln)}$: is the set of
        verifiable \vuln\ which responsible code is not detected in version
        \version.
    \item $\unverifiable(\version) = \Set{\vuln| \version \in
        \Vreport(\vuln) \wedge \Vverified(\vuln) = \bot}$: is the set of
        unverifiable \vuln\ of version \version.
\end{itemize}
For example, according to \tabref{tbl:method:example},
\Vreport(\cve{2011-4080}) = \Set{\ver{1}-\ver{8}}, \Vverified(\cve{2011-4080})
= \Set{\ver{3}-\ver{8}}. Then, \cve{2011-4080} is a verified \cve\ in \ver{3}
\ie $\cve{2011-4080} \in \verified(\ver{3})$; whereas it is an erroneous \cve\
in \ver{1} \ie $\cve{2011-4080} \in \erroneous(\ver{1})$.

By ignoring the unverifiable vulnerabilities, the error rate of a version
\version\ of Chrome is defined as the ratio of the number of erroneous
vulnerabilities of \version\ by the number of verifiable vulnerabilities of
\version, as shown in the following formula:
\begin{equation}
    \textit{ER}(\version) = \frac{|\erroneous(\version)|}{|\verified(\version)| + |\erroneous(\version)|} \label{eq:error-rate}
\end{equation}

Being more optimistic, we assume all unverifiable vulnerabilities are all
correct. The error rate is rewritten as follows:
\begin{equation}
    \textit{ER}'(\version) = \frac{|\erroneous(\version)|}
        {|\unverifiable(\version)| + |\verified(\version)| + |\erroneous(\version)|}
\end{equation}

\begin{figure}[t]
    \centering
    \subfigure[Error Rate]{
        \includegraphics[width=0.45\columnwidth,height=6\baselineskip]{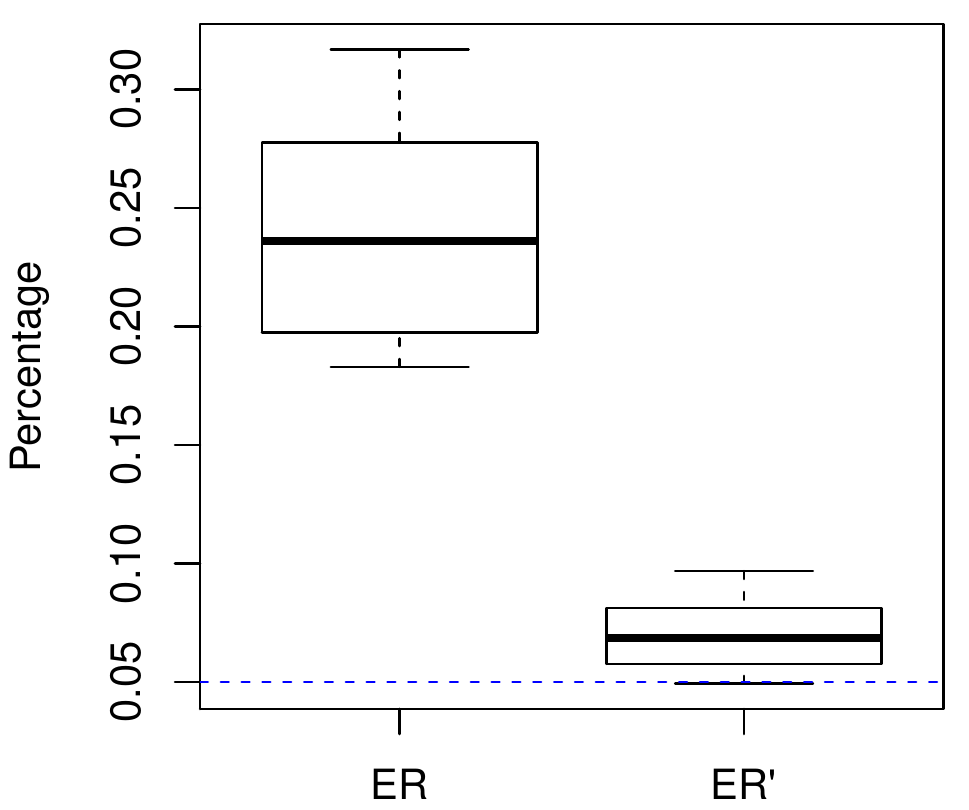}
        \label{fig:vuln:error}
    }
    \subfigure[Types of Error] {
        \includegraphics[width=0.45\columnwidth,height=6\baselineskip]{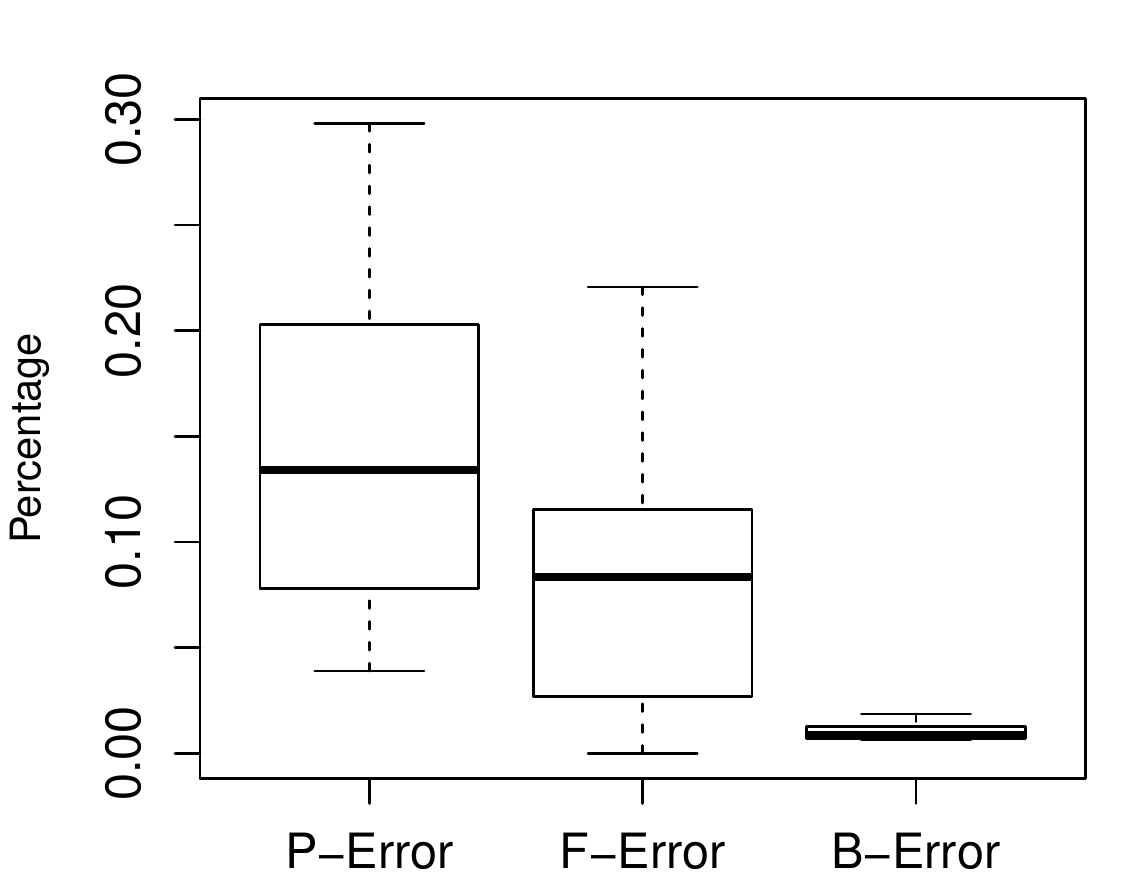}
        \label{fig:error:err-type:boxplot}
    }
    \ACCORCIA
    \extcaption[\columnwidth]{In the whicker-box plot, the whickers represent
    the min and max value, the bold line in the middle is the median value, and the
    lower and upper part of the box are the quartile of the distribution. The blue dash line at $0.05$
    shows the threshold of error rate.}
    \ACCORCIA
    \caption{The errors in vulnerable version data of NVD entries for Chrome.}
    \ACCORCIA
\end{figure}

\figref{fig:vuln:error} shows the box plots for the distribution of the error
rates in Chrome versions. In a box plot, the whickers represent the min and max
values, the bold line in the middle is the median, and the lower and upper
parts of the box are the quartiles of the distribution. According to the
figure, since the median of both error rates $ER, ER'$ are much greater than
\Threshold. It remarkably denotes that the number of erroneous vulnerabilities
is not negligible. This is confirmed in the one-sided Wilcoxon rank-sign test
where the null hypothesis is \emph{``the median of error rates is \Threshold"},
and the alternative hypothesis is \ref{hypo:significant:err-rate}. The returned
\pvalue\ for $ER$ and $ER'$ are almost zero ($2.44 \cdot 10^{-4}$, and $1.22
\cdot 10^{-4}$respectively). It means the error rates (both $ER$ and $ER'$) did
not randomly happen and therefore are significantly greater than \Threshold.

We break down erroneous vulnerabilities into following categories:
\begin{itemize}
    \item \emph{stretched-past error} (\PErr): is the set of erroneous
        vulnerabilities whose version \version\ is older than all versions
        that the NVD entries are verified to impact to.
        \[
            \PErr(\version) = \Set{\vuln \in \erroneous(\version)| \version < \min(\Vverified(\vuln))}
        \]
    \item \emph{future-version error} (\FErr): is the set of erroneous
        vulnerabilities whose version \version\ is newer than all versions
        that the NVD entries are verified to impact to.
        \[
            \FErr(\version) = \Set{\vuln \in \erroneous(\version)| \version > \max(\Vverified(\vuln))}
        \]
    \item \emph{beta error} (\BErr): is the set of erroneous
        vulnerabilities whose corresponding NVD entries only impact
        non-official versions, \ie $\Vverified(\vuln) = \emptyset$.
        \[
            \BErr(\version) = \Set{\vuln \in \erroneous(\version)| \Vverified(\vuln) = \emptyset}
        \]
\end{itemize}

Similarly to \eqref{eq:error-rate}, we calculate the \emph{stretched-past error
rate, future-error rate}, and \emph{beta error rate}.
\figref{fig:error:err-type:boxplot} reports the distributions of these rates.
The \EErr\ is slightly greater than \LErr, and both of them are much greater
than \BErr. \BErr\ seems to be negligible. We employ Wilcoxon rank-sum test to
compare each pair of error categories. Since we compare one category to other
two, the Bonferroni correction is applied \ie the significance level is divided
by 2: $\alpha =^{0.05}/_2 = 0.025$. The test result confirms that both \EErr\
and \LErr\ are significantly greater than \BErr\ since the returned \pvalue s
are less than $\alpha$. The  $\pvalue = 0.03 > \alpha$ of the comparison
between \PErr\ and \FErr\ can be considered an evidence (even if not
significant) that \PErr\ is greater than \FErr.


\begin{figure}
    \centering
    \includegraphics[width=1\columnwidth]{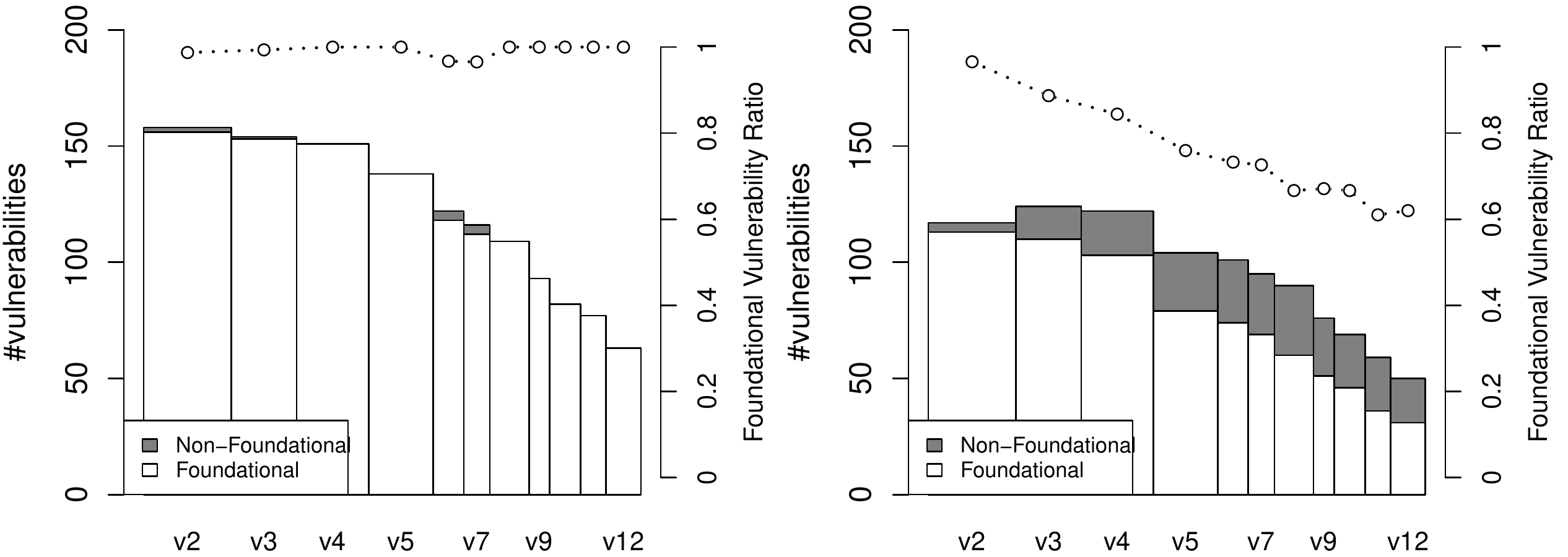}
    \ACCORCIA[-1.5]
    \caption{Verifiable (left) vs. Verified (right) vulnerabilities.}
    \label{fig:vulns:verification}
    \ACCORCIA
\end{figure}

\begin{figure}
    \centering
    \subfigure[Verifiable vulnerabilities] {
        \includegraphics[width=\columnwidth]{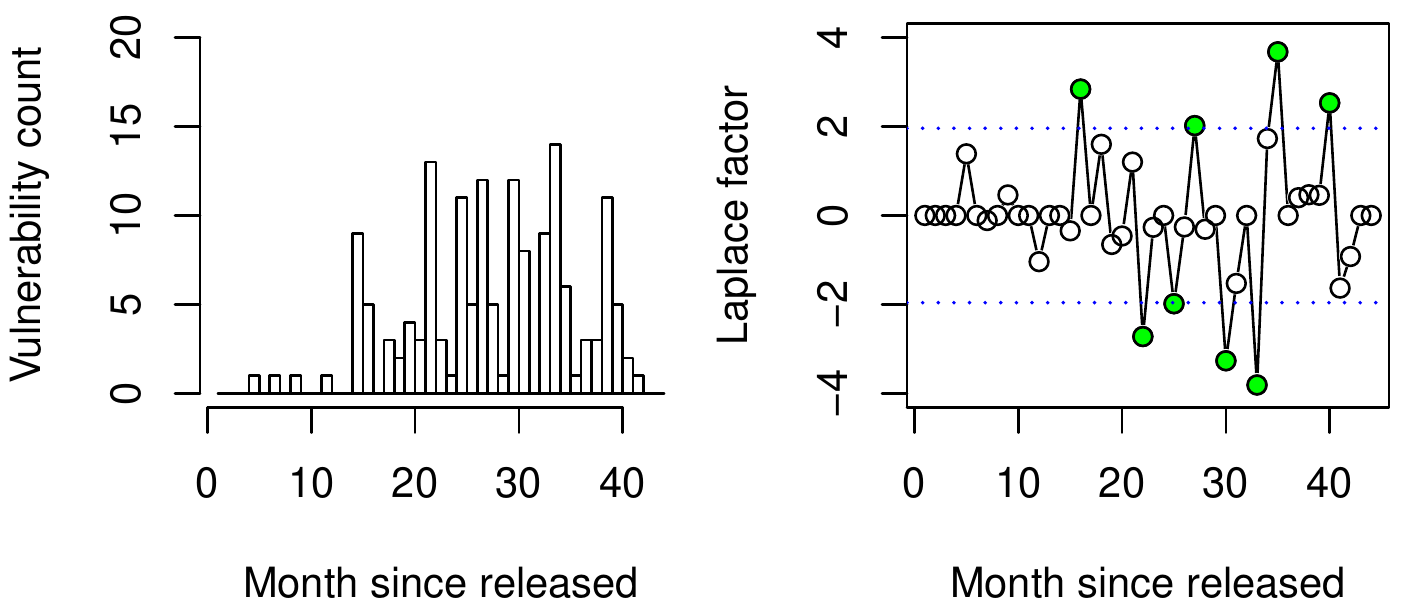}
        \label{fig:trend:verifiable}
    }
    \subfigure[Verified vulnerabilities] {
        \includegraphics[width=\columnwidth]{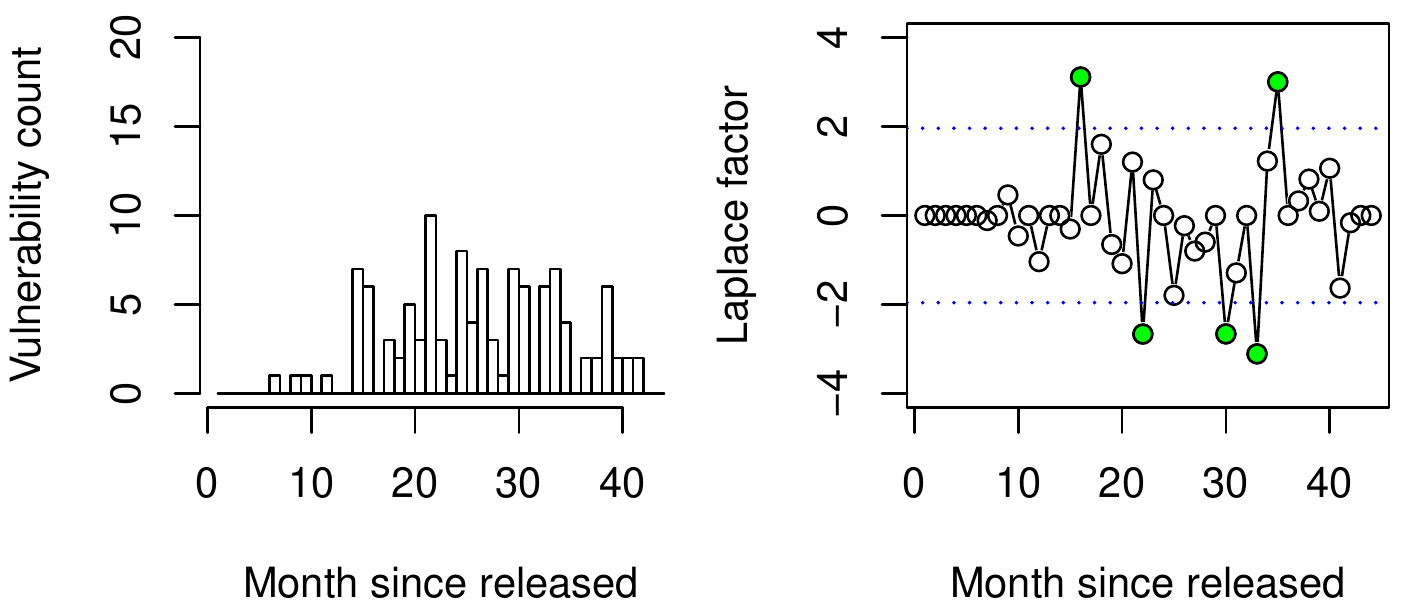}
        \label{fig:trend:verified}
    }
    \ACCORCIA
    \caption{Trend of foundational vulnerability discovery.}
    \label{fig:trend}
    \ACCORCIA
\end{figure}

Hereafter we revise the running example about foundational vulnerability in
Chrome. We assume that the same ratio of errors would be applied in the
unverifiable vulnerabilities. Therefore, in following analysis we study the
impact of error in verifiable vulnerabilities when we study the trend of
foundational vulnerability.

We have two data sets: \dsVerifiable\ and \dsVerified. The former is the set of
verifiable vulnerabilities that we could verify by the proposed method. Their
vulnerable versions are reported by NVD. The latter is also the same set, but
their vulnerable versions are verified.

\figref{fig:vulns:verification} illustrates the fraction of foundational
vulnerabilities in each version. Left is based on \dsVerifiable, and right is
based on \dsVerified. In this figure, The circles denote the percentage of
foundational vulnerabilities. By looking at the left picture, even though the
absolute number of foundational vulnerabilities decreased, their fractions are
almost unchanged. Additionally, as aforementioned, it is strange that
vulnerabilities are only introduced in \ver{1.0}, but none are introduced in
later versions. However, by looking at the right side, this phenomenon
disappears. Moreover, there is a decreasing trend in the foundational
vulnerabilities fractions from \ver{2.0} to \ver{12.0}. We additionally employ
the Wilcoxon rank-sum test on the absolute number and the fraction of
foundational vulnerabilities in each version. The test results show that the
difference between the left and the right is indeed not random since the
returned \pvalue s are almost zero (\ie $3.82\cdot 10^{-3}$, and $3.84\cdot
10^{-3}$ respectively).

Furthermore, we replicate the analysis on the trend of foundational
vulnerability discovery as described in \cite{OZMEN-SCHE-06-USENIX}.
\figref{fig:trend} exhibits the analysis result on \dsVerifiable\
(\figref{fig:trend:verifiable}) and \dsVerified (\figref{fig:trend:verified}).
In the figure, left is the discovery rate of foundational vulnerabilities
discovered monthly since the release date, right is the Laplace test for trend
in monthly discovered foundational vulnerabilities. Two dotted horizonal lines
at value $1.96$ and $-1.96$ indicate the range such that if a value of Laplace
factor is out of this range, it is significant evidence for either an
increasing ($>1.96$) or a decreasing ($<-1.96$) trend in the data. Such values
are indicated as green (gray) circles. Again we see a clearly difference
between two discovery rates between two data sets. The \pvalue\ of the Wilcoxon
rank-sum test is almost zero ($0.0008$): this indicates that the difference is
significant. We can also see the difference in the trend of discovery. By using
\dsVerifiable, we might observe several significant trends (both increasing and
decreasing) of foundational vulnerability discovery. Some of these trends,
however, disappear in \dsVerified.

In short, our experiment provides evidence that the error in the
\emph{vulnerable versions} feature of NVD entries for Chrome is not negligible.
Among the errors, NVD tends to commit more stretch-past error than others. It
is one of the reasons for the abnormality that $99.5\%$ vulnerabilities of
Chrome are foundational. This error in NVD has significantly impact to the
analysis of foundational vulnerabilities where different conclusions can be
drawn.

\section{Threat to Validity}\label{sec:threat}
\noindent\textbf{Construct validity} includes threats affecting the approach by
means of which we collect and verify vulnerabilities. Threats in this category
come from the assumptions as follows.

By making the assumption \ref{ass:commit:message}, we delegate the completeness
of our method to the responsibility of developers and the quality control of
the software vendor. According to \cite{BIRD-etal-09-ESEC-FSE}, there are two
types of mistakes: the developers do not mention the bug ID in a bug-fix
commit; and the developers mention a bug ID in a non-bug-fix commit. Also in
\cite{BIRD-etal-09-ESEC-FSE}, the authors showed that the latter is negligible,
while the former does exist. To evaluate the impact of the latter mistakes, we
have done a qualitative analysis on bug-fix reports, and we found that all
analyzed bug-fix commits are actually bug fixes. As for the former mistakes, we
check the completeness of the bug-fix commits for vulnerabilities. As
discussed, we found a large portion of vulnerabilities for which we could not
locate the bug-fix commits. Our qualitative analysis on these vulnerabilities
reveals that they originate from external projects used in Chrome. We discuss a
potential solution addressing this threat in future work.

The second assumption \ref{ass:responsible:code} is apparently syntactical and
might not cover all the cases of bug fixes since it is extremely hard to
automatically understand the root of vulnerabilities. The assumption also means
that if a version contains at least one line of responsible code, this version
is vulnerable. Together with the assumption \ref{ass:responsible:missing}, our
method might overestimate the vulnerable versions by classifying safety code as
buggy (error type I, false positive). However, since most Chrome
vulnerabilities are reported as foundational, if we overestimate the vulnerable
version, the reported \emph{error rate} is a lower bound of the actual one.

Besides, a technical threat to construct validity may be the buggy
implementation of the method. To minimize such problem, we employ multi-round
test-and-fix approach where we ran the program on some vulnerabilities, then we
manually checked the output, and fixed found bugs. We repeated this procedure
until no bug has been found. Finally, we randomly checked the output again to
ensure there was no mistake.

\noindent {\bf Internal validity} concerns the causal relationship between the
collected data and the conclusion withdrawn in our study. Our conclusions are
based on statistical tests. These tests have their own assumptions. Choosing
tests whose assumptions are violated might end up with wrong conclusions. To
reduce the risk we carefully analyzed the assumptions of the tests: for
instance, we did not apply any tests with normality assumption since the
distribution of vulnerabilities is not normal.

\noindent {\bf External validity} is the extent to which our conclusion could
be generalized to other applications. Our experiment is based on the
vulnerability data of Chrome. So, to have a more generalized conclusion, a
replication of this work on other applications should be done.

\section{Related Work} \label{sec:relatedwork}
%

\'Sliwerski et al\cite{SLIW-05-MSR} proposed a technique that automatically
locates fix-inducing changes. This technique first locates changes for bug
fixes in the commit log, then determines earlier changes at these locations.
These earlier changes are considered as the cause of the later fixes, and are
called fix-inducing. This technique has been employed in several studies
\cite{SLIW-05-MSR,ZIMM-etal-07-PROMISE} to construct bug-fix data sets.
However, none of these studies mention how to address bug fixes which earlier
changes could not be determined. These bug fixes were ignored and became a
source of bias in their work.

Bird et al\cite{BIRD-etal-09-ESEC-FSE} conducted a study the level bias of
techniques to locate bug fixes in code base. The authors have gathered a data
set linking bugs and fixes in code base for five open source projects, and
manual checked for the biases in their data set. They have found strong
evidence of systematic bias in bug-fixes in their data set. Such bias might be
also existed in other bug-fix data set, and could be a critical problem to any
study relied on such biased data.

Antoniol et al\cite{ANTONIOL-etal-08-CASCON} showed another kind of bias that
the bug-fixes data set might suffer from. Many issues reported in many tracking
system are not actual bug reports, but feature or improvement requests.
Therefore, this might lead to inaccurate bug counts. However, such bias rarely
happens for security bug reports. Furthermore, Nguyen et
al\cite{NGUY-ADAM-HASS-10-WCRE}, in an empirical study about bug-fix data sets,
showed that the bias in linking bugs and fixes is the symptom of the software
development process, not the issue of the used technique. Additionally, the
linking bias has a stronger effect than the \emph{bug-report-is-not-a-bug}
bias.

\section{Conclusion and Future Work} \label{sec:conclusion}
In this paper we have conducted an experiment to verify the reliability of the
\emph{vulnerable versions} data of Chrome vulnerabilities reported by NVD. The
experiment has revealed that the error in the \emph{vulnerable versions} data
is notable. Among verifiable vulnerabilities of individual Chrome version,
approximately $25\%$ of them are erroneous. If we assume that all unverifiable
vulnerabilities are all correct, still more than $7\%$ are erroneous overall.
We also demonstrated how these erroneous vulnerabilities could potentially
impact the conclusion of foundational vulnerability study. Another study on the
impact of erroneous vulnerabilities is further discussed in
\techreport[\cite{NGUY-MASS-CCR-XXX}]{Appendix \ref{app:vdm}}. This experiment
has shed a light into the (un)reliability of the NVD, and allows researchers to
revisit the reliability of existing vulnerability databases.

However about two-third of Chrome vulnerabilities are unverifiable because they
are vulnerabilities of the external projects used in Chrome. To be able to
verify them, extra effort is required. First, we need to link the unverifiable
vulnerabilities to the bug ID of the external projects. This could be done by
parsing the Chrome bug report. Our qualitative study on several unverifiable
vulnerability reports shows that all of them have links to bug reports of the
external projects. Second, we apply the proposed method to identify vulnerable
revisions of the external projects. Finally, we link these vulnerable revisions
to the version of Chrome by looking at the repository of Chrome. For example,
Chrome \ver{12.0} uses WebKit revision \WebKetRevision, V8 revision
\JSErevision. A more detail discussion can found in
\techreport[\cite{NGUY-MASS-CCR-XXX}]{Appendix \ref{app:externalprj}}.

Also as a part of future work, we plan to evaluate the robustness of the
proposed method in identifying vulnerable revisions correctly. We also plan to
repeat the experiment on other open source software like Firefox to have a
better insight about the reliability of NVD.

\bibliographystyle{abbrv}
\techreport[\bibliography{../../Biblio/short-names,../../Biblio/security-common,datasources}]{}

\iftechreport

\appendix
\section{Vulnerability Discovery Model Revisit}\label{app:vdm}
This section analyzes to what extent the erroneous vulnerabilities can affect
the conclusions in the validation experiment on vulnerability discovery models
(VDM) \cite{NGUY-MASS-12-ASIACCS}. Readers are referred to
\cite{NGUY-MASS-12-ASIACCS} for full detail of the experiment as well as the
metrics. Here we only sketch the experiment and its conclusions that might be
affected.

\subsection{Validation Experiment Overview} \label{sec:vdm:review} Nguyen and
Massacci\cite{NGUY-MASS-12-ASIACCS} conducted an experiment to study the
performance of six VDM (\ie AML, AT, LN, LP, RE, and RQ\footnote{The full names
of these models are: AML - Alhazmi-Malaiya Logistic; AT - Anderson
Thermodynamic; LN - Linear; LP - Logistic Poison; RE - Rescorla Exponential
(RE); RQ - Rescorla Quadratic.}) on six versions of Chrome (\ver{1}--\ver{6})
and other browsers. That experiment fitted all VDM to the monthly-observed
vulnerabilities of each version from the sixth month after released to present
(time of writing). The goodness-of-fit of VDM is assessed $\chi^2$
goodness-of-fit tests. Based on the returned \emph{p-values} of these tests,
the authors calculated two quantitative metrics, namely \emph{goodness-of-fit
entropy $E_\beta(t)$}, and \emph{goodness-of-fit quality $Q_\omega(t)$}. These
metrics are used to analyze stability of the data sets and the overall
performance of VDM are analyzed.  Therefore, to study the impact of the
erroneous vulnerabilities, we replicate the experiments on two data sets of
verifiable vulnerabilities as below:
\begin{align*}
    \texttt{nvd.verifiable} &= \Set{\Seq{\vuln,\version}|\Vverified(\vuln) \not=\bot} \\
    \texttt{nvd.verified} &= \Set{\Seq{\vuln, \version}|\version \in \Vverified(\vuln)}
\end{align*}

\subsection{Experiment Revisit}
We fit all VDM to the monthly-observed vulnerabilities of each of fourteen
versions of Chrome \ver{1}--\ver{12} from the sixth month after released to
present\footnote{the time of data collection, July 2012}. Then we calculate the
entropy and quality of each VDM in each dataset.

\figref{fig:entropy:line} plots the evolution of entropy for both
\texttt{nvd.verifiable} (solid line) and \texttt{nvd.verified} (dash line).
\figref{fig:entropy:box} shows the distribution of the entropy of these two
data sets. The entropy of \texttt{nvd.verifiable} seems to be a bit greater
than that of \texttt{nvd.verified}. We perform a paired Wilcoxon test to check
the null hypothesis that \emph{``there is no different between the medians of
the entropies of the two data sets}. The returned \emph{p-value} of $0.04$
shows that the difference between the entropies of the two data sets is
statistically significant.

\begin{figure}[t]
    \centering
    \subfigure[]{
        \includegraphics[width=0.45\textwidth]{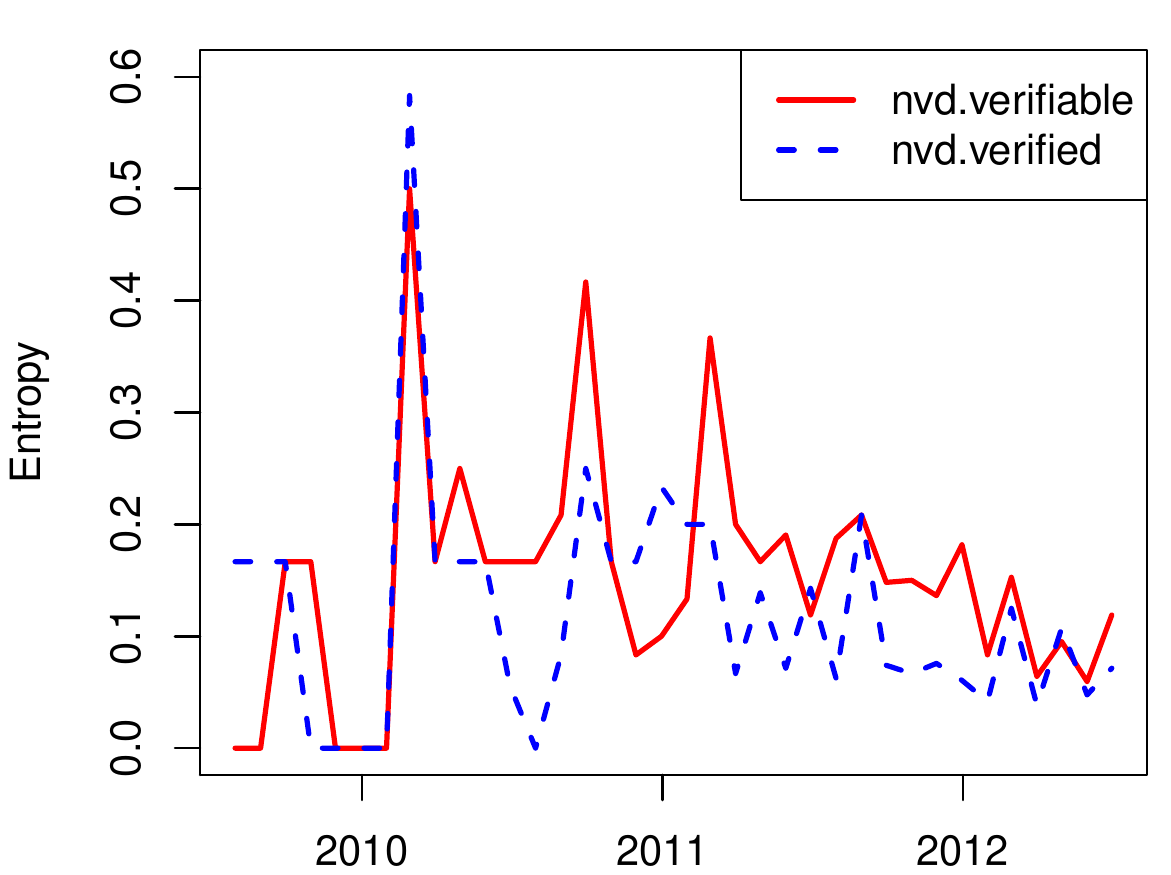}
        \label{fig:entropy:line}}
    \subfigure[]{
        \includegraphics[width=0.45\textwidth]{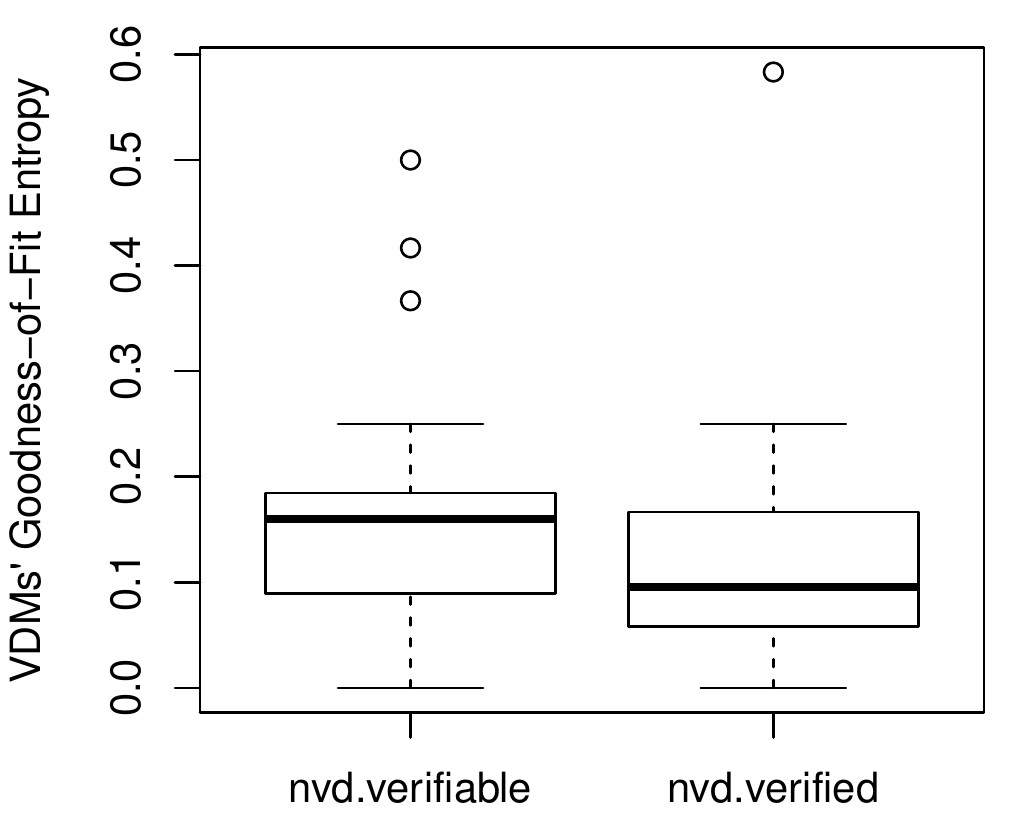}
        \label{fig:entropy:box}}
    \ACCORCIA
    \caption{The entropy $E_1(t)$ between \texttt{nvd.verifiable} and \texttt{nvd.verified}.}
    \label{fig:entropy}
    \ACCORCIA
\end{figure}

\begin{figure*}
    \centering
    \includegraphics[width=0.9\textwidth]{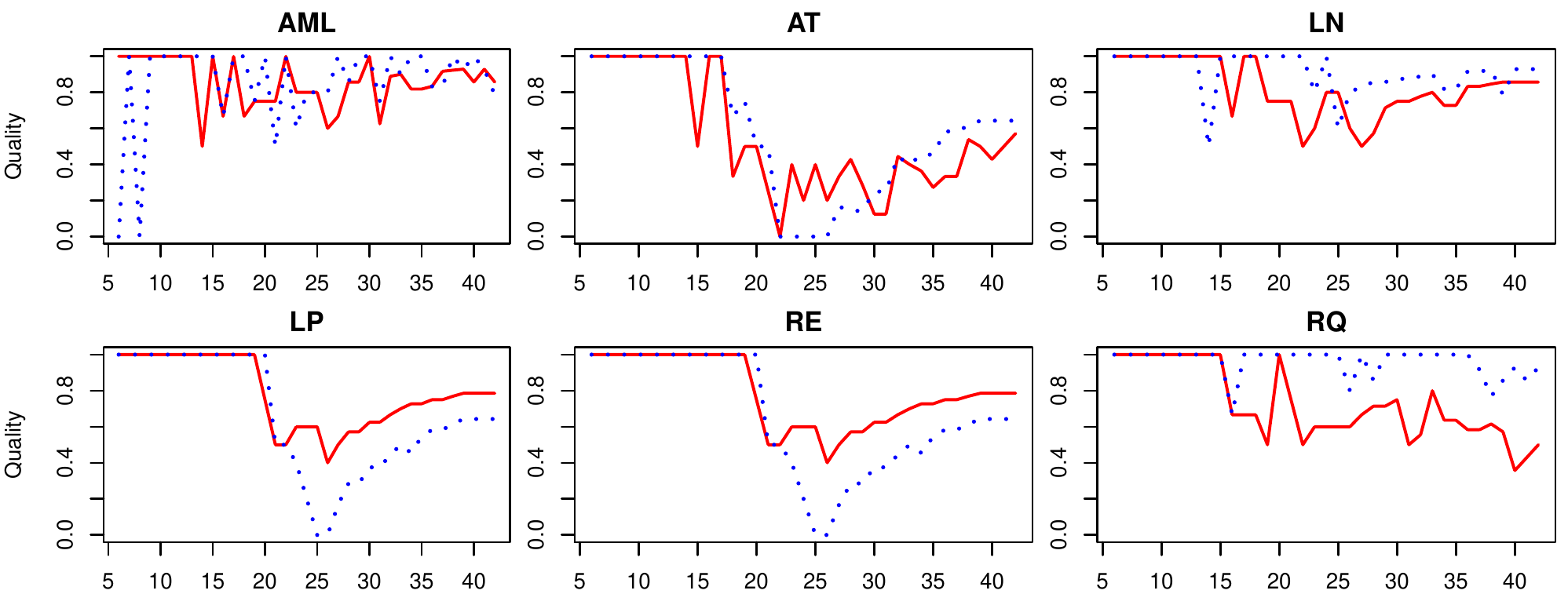}
    \extcaption[0.9\textwidth]{This figure plots the quality of VDM in the two data sets
    \texttt{nvd.verifiable} (solid lines) and \texttt{nvd.verified} (dotted lines).
    The X-axis is the month since release, and the Y-axis is the quality of VDM, \ie the ratio
    of the number of times the VDM well fits the observed data ($\pvalue \ge 0.05$) by the total
    number of time the VDM is fitted to the observed data.}
    \caption{The quality of VDM in \texttt{nvd.verifiable} (solid lines) and \texttt{nvd.verified} (dotted lines).}
    \label{fig:quality:line}
\end{figure*}

\figref{fig:quality:line} plots the quality of the VDM in the two data sets
\texttt{nvd.verifiable} (solid lines) and \texttt{nvd.verified} (dotted lines).
The X-axis is the month since release (MSR), and the Y-axis is the quality of
VDM, \ie the ratio of the number of times the VDM well fits the observed data
($\pvalue \ge 0.05$) by the total number of time the VDM is fitted to the
observed data. The qualities of the five models AT, LN, LP, RE, and RQ between
the two data sets are mostly the same in the some first MSR. But these
qualities are much different in later MSR. Meanwhile, the qualities of the AML
are very different in some first MSR, but get closer in later MSR.

We perform the paired Wilcoxon test for the qualities of VDM. The null
hypothesis is \emph{``there is no different between the quality of VDM in two
data sets"}. The returned $\pvalue = 0.15 > 0.05$ is a weak evidence about the
difference between qualities in the two data sets. We additionally perform the
paired Wilcoxon test for the qualities of VDM individually in two data sets.
The null hypothesis is also \emph{``there is no different between the quality
of VDM in two data sets"}. \tabref{tbl:quality:vdm} reports the \pvalue s of
the tests. The $\pvalue < 0.05$ (bold value) denotes the significance of the
difference between the qualities of VDM in two data sets. According to the
table, the qualities of four models LN, LP, RE and RQ in two data sets are
significantly different; whereas the qualities of AML and AT in two data sets
are not able to conclude.

\begin{table}[t]
    \centering
    \caption{The \emph{p-values} of the tests of VDM quality in two data sets.}
    \label{tbl:quality:vdm}
    \extcaption{The table reports the \pvalue s of the tests of the null hypothesis
    that \emph{``there is no different between the quality of VDM in two data sets"}.
    The bold values less than 0.05 denote the significant of the test.}
    \begin{tabular}{ccccccc}
        \toprule
        & AML & AT & LN & LP & RE & RQ \\
        \midrule
        \emph{p-value} & $0.32$ & $0.35$ & $\bf <0.01$ &  $\bf <0.01$ & $\bf <0.01$ & $\bf < 0.01$ \\
        \bottomrule
    \end{tabular}
\end{table}

In summary, the erroneous vulnerabilities significantly change the entropy of
individual VDM. For the quality, the impact of the erroneous vulnerabilities to
the individual VDM's performance is different. The error does not change much
the performance of AML, AT; whereas, it does significantly change the
performance of LN, LP, RE and RQ model. However, we only have a weak evidence
about the difference of the overall quality of VDM in the two data sets. and so
does the entropy. Therefore, erroneous vulnerabilities of Chrome partially
change the performance of VDM as well as the conclusions in
\cite{NGUY-MASS-12-ASIACCS}.

\section{Dealing with Bug Fixes in External Projects}\label{app:externalprj}
\fi

\end{document}